\documentclass[aps,pra,reprint,showpacs,floatfix,longbibliography]{revtex4-1}
\usepackage{graphicx, amsmath, amssymb, hyperref}
\usepackage[caption=false]{subfig}

\begin{document}

%-------------------------------------------------------------------------------
% Shortcut Commands
%-------------------------------------------------------------------------------

\newcommand{\braket}[2]{{\left\langle #1 \middle| #2 \right\rangle}}
\newcommand{\bra}[1]{{\left\langle #1 \right|}}
\newcommand{\ket}[1]{{\left| #1 \right\rangle}}
\newcommand{\ketbra}[2]{{\left| #1 \middle\rangle \middle \langle #2 \right|}}
\newtheorem{theorem}{Theorem}

%-------------------------------------------------------------------------------
% Front Matter
%-------------------------------------------------------------------------------

\title{Optimal Quantum Walk Search on Kronecker Graphs \\ with Dominant or Fixed Regular Initiators}

\author{Adam Glos}
	\email{aglos@iitis.pl}
	\thanks{Authors in alphabetical order}
	\affiliation{Institute of Theoretical and Applied Informatics, Polish Academy of Sciences, Ba{\l}tycka 5, 44-100 Gliwice, Poland}
	\affiliation{Institute of Informatics, Silesian University of Technology, ul. Akademicka 16, 44-100 Gliwice, Poland}
\author{Thomas G.~Wong}
	\email{thomaswong@creighton.edu}
	\affiliation{Department of Physics, Creighton University, 2500 California Plaza, Omaha, NE 68178, USA}

\begin{abstract}
	In network science, graphs obtained by taking the Kronecker or tensor power of the adjacency matrix of an initiator graph are used to construct complex networks. In this paper, we analytically prove sufficient conditions under which such Kronecker graphs can be searched by a continuous-time quantum walk in optimal $\Theta(\sqrt{N})$ time. First, if the initiator is regular and its adjacency matrix has a dominant principal eigenvalue, meaning its unique largest eigenvalue asymptotically dominates the other eigenvalues in magnitude, then the Kronecker graphs generated by this initiator can be quantum searched with probability $1$ in $\pi\sqrt{N}/2$ time, asymptotically, and we give the critical jumping rate of the walk that enables this. Second, for any fixed initiator that is regular, non\nobreakdash-bipartite, and connected, the Kronecker graphs generated by it are quantum searched in $\Theta(\sqrt{N})$ time. This greatly extends the number of Kronecker graphs on which quantum walks are known to optimally search. If the fixed, regular, connected initiator is bipartite, however, then search on its Kronecker powers is not optimal, but is still better than classical computer's $O(N)$ runtime if the initiator has more than two vertices.
\end{abstract}

\pacs{03.67.Ac, 03.67.Lx}

\maketitle

%-------------------------------------------------------------------------------
% Main Matter
%-------------------------------------------------------------------------------

\section{Introduction}

Many real-world networks, despite occurring in vastly different physical systems ranging from molecular interactions in cells \cite{Barabasi2004} to computer networks \cite{snapnets}, share similar properties. For example, networks tend to be small-world \cite{Milgram1967}, meaning the number of hops to reach any node from another is small. Real networks are also often scale-free \cite{Barabasi1999}, meaning the distribution of the number of neighbors of each node is heavy-tailed or follows a power law. Several models have been proposed to generate networks possessing such properties \cite{Barabasi1999,WS1998,Ravasz2002}. One model is to use Kronecker graphs \cite{Leskovec2005a}, and they have been used to generate graphs that mimic the network of citations of arXiv preprints and U.S.~patents, and the trust network of the Epinions social network \cite{Leskovec2010}.

\begin{figure}
\begin{center}
	\includegraphics{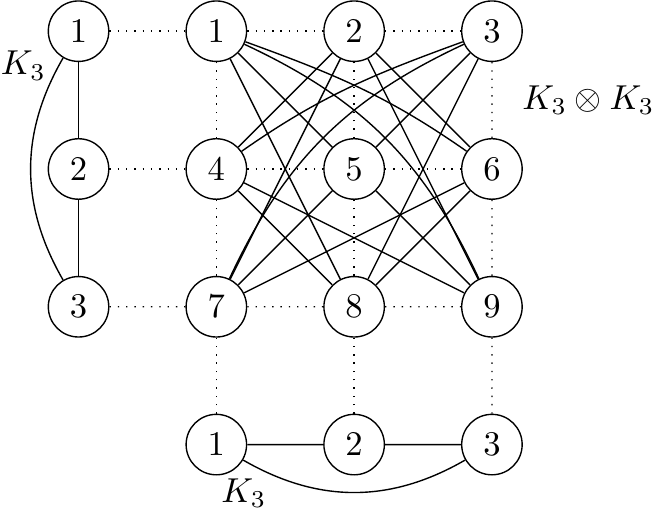}
	\caption{The complete graph of three vertices $K_3$ and its second-order Kronecker power $K_3 \otimes K_3$.}\label{fig:K3K3}
\end{center}
\end{figure}

In the deterministic model of Kronecker graphs, one begins with an ``initiator'' graph of $M$ vertices whose adjacency matrix $A$ is an $M \times M$ matrix, where $A_{ij} = 1$ if vertices $i$ and $j$ are adjacent and $0$ otherwise. The $j$th order Kronecker graph is the graph whose adjacency matrix is
\begin{equation}
	\label{eq:def}
	A^{\otimes j} = \underbrace{A \otimes A \otimes \dots \otimes A}_{j{\rm~times}},
\end{equation}
where $\otimes$ denotes the Kronecker or tensor product. The resulting Kronecker graph has $N = M^j$ vertices. For example, the adjacency matrix of the complete graph of three vertices is
\[ A = \left( \begin{array}{ccc}
	0 & 1 & 1 \\
	1 & 0 & 1 \\
	1 & 1 & 0 \\
\end{array} \right). \]
Then the second-order Kronecker graph generated by this has $3^2 = 9$ vertices, and its adjacency matrix is
\[ A^{\otimes 2} = A \otimes A = \left( \begin{array}{ccccccccc}
	0 & 0 & 0 & 0 & 1 & 1 & 0 & 1 & 1 \\
	0 & 0 & 0 & 1 & 0 & 1 & 1 & 0 & 1 \\
	0 & 0 & 0 & 1 & 1 & 0 & 1 & 1 & 0 \\
	0 & 1 & 1 & 0 & 0 & 0 & 0 & 1 & 1 \\
	1 & 0 & 1 & 0 & 0 & 0 & 1 & 0 & 1 \\
	1 & 1 & 0 & 0 & 0 & 0 & 1 & 1 & 0 \\
	0 & 1 & 1 & 0 & 1 & 1 & 0 & 0 & 0 \\
	1 & 0 & 1 & 1 & 0 & 1 & 0 & 0 & 0 \\
	1 & 1 & 0 & 1 & 1 & 0 & 0 & 0 & 0 \\
\end{array} \right). \]
The resulting graph is shown in Fig.~\ref{fig:K3K3}. 

Recently, Wong \textit{et al.}~\cite{Wong29} proposed investigating how quickly a quantum computer searches Kronecker graphs for a marked node using a quantum walk. Quantum walks are a model of universal quantum computation \cite{Childs2009} that have been used to develop quantum algorithms for searching \cite{SKW2003}, element distinctness \cite{Ambainis2004}, triangle finding \cite{MSS2005}, and evaluating boolean formulas \cite{FGG2008}. Searching the complete graph of $N$ vertices is equivalent to searching an unordered database of $N$ items, and a quantum walk accomplishes this in $O(\sqrt{N})$ time \cite{AKR2005,CG2004}, the same as Grover's algorithm \cite{Grover1996}. If the graph is incomplete, however, it is generally unknown under what conditions a graph supports optimal quantum search \cite{BBBV1997}, i.e., in $O(\sqrt{N})$ time, although global symmetry \cite{Wong5}, connectivity \cite{Wong7}, and regularity \cite{Novo2015} are some properties that have been explored.

In this paper, we consider search on Kronecker graphs using a continuous-time quantum walk \cite{CG2004}, where the quantum walk is effected by the adjacency matrix. For a regular graph, this is equivalent to a quantum walk effected by the graph Laplacian \cite{CG2004}, but if the graph is irregular, the two quantum walks can differ \cite{Wong19}. The system $\ket{\psi(t)}$ begins in a uniform superposition $\ket{s}$ over all $N$ vertices: 
\begin{equation}
	\label{eq:psi0}
	\ket{\psi(0)} = \ket{s} = \frac{1}{\sqrt{N}} \sum_{i=1}^N \ket{i}.
\end{equation}
In the adjacency quantum walk, this evolves by Schr\"odinger's equation with Hamiltonian
\begin{equation}
	\label{eq:H}
	H = -\gamma A^{\otimes j} - \ketbra{w}{w},
\end{equation}
where $\gamma$ is a positive, real parameter corresponding to the jumping rate (amplitude per unit time) of the quantum walk, and $\ket{w}$ is the marked vertex to search for. Computationally, the first term effects the quantum walk, and the second term is a Hamiltonian oracle \cite{Mochon2007}. The jumping rate must be judiciously chosen to take some ``critical value'' $\gamma_c$ in order for the system to evolve beyond a trivial phase \cite{CG2004}.

Mathematically, one can instead consider a Hamiltonian with positive terms:
\begin{equation}
	\label{eq:Hplus}
	H^+ = \gamma A^{\otimes j} + \ketbra{w}{w}.
\end{equation}
Although evolving by $H$ and $H^+$ for time $t$ results in different quantum states, the probability of getting $\ket{w}$ when measuring the position of the walker (i.e., the success probability) is identical in both cases. Physically, evolving by $H^+$ for time $t$ is equivalent to evolving by $H$ for time $-t$, i.e., backward in time.

Wong \textit{et al.}~\cite{Wong29} explored this search algorithm on Kronecker graphs where the initiator was the complete graph of $M$ vertices. They completely solved it for Kronecker powers 1, 2, and 3, giving the critical jumping rate $\gamma_c$  and proving that the success probability reaches $1$ at time $\pi\sqrt{N}/2$, asymptotically. They also conjectured from numerical simulations that higher-order Kronecker graphs with the complete initiator are also searched in the same time, but their analytical method of degenerate perturbation theory was not conducive to proving it analytically.

In this paper, we generalize this by considering Kronecker graphs where the initiator is regular and has the property that its adjacency matrix $A$ has a unique principal eigenvalue $\lambda_{A,1}$ that dominates in magnitude the other eigenvalues $\lambda_{A,2}, \dots, \lambda_{A,M}$, asymptotically for large $M$. That is, using little-$o$ notation \cite{Knuth1976}, for every $i = 2, 3, \dots, M$ we have $\lambda_{A,i} = o(\lambda_{A,1})$.  This includes the complete graph of $M$ vertices as a special case, since its adjacency eigenvalues are
\begin{equation}
	\label{eq:complete}
\lambda_{A,1} = M-1 \quad \text{and} \quad \lambda_{A,i \ge 2} = -1,
\end{equation}
with respective multiplicities $1$ and $M-1$, so the principal eigenvalue is unique and dominates all the others in magnitude for large $M$. So when we say an initiator has a dominant principal eigenvalue, we mean a sequence of initiators where $\lambda_{A,1}$ is increasingly dominant as $M$ increases. Note that possessing a dominant principal eigenvalue is stronger than possessing a spectral gap, since a spectral graph only assumes that the principal eigenvalue dominates the second eigenvalue. For example, as we will discuss in Sec.~\ref{sec:shift}, the regular complete bipartite graph has a spectral gap, but its principal eigenvalue does not dominate all the other eigenvalues.

In Sec.~\ref{sec:dominant}, we prove that all Kronecker graphs generated by such regular, dominant-eigenvalue initiators asymptotically support optimal quantum search, reaching a success probability of $1$ at time $\pi\sqrt{N}/2$, and we give the critical jumping rate $\gamma_c$ that enables this. We prove this using properties of Kronecker products, a Lemma by Chakraborty \textit{et al.}~\cite{Chakraborty2016}, and an extension by Glos \textit{et al.}~\cite{Glos2018}. This general result proves Wong \textit{et al.}'s conjecture with the complete initiator \cite{Wong29} as a special case. Then in Sec.~\ref{sec:shift}, we explore graph connectivity and optimal quantum search, showing that shifting and rescaling the quantum walk term of the Hamiltonian is necessary in some situations for proving the optimality of quantum search, and we give optimal parameters that maximize the lower bound on the success probability. Finally, in Sec.~\ref{sec:fixed}, we explore fixed initiators (where $M$ is constant) and prove that if they are regular, non-bipartite, and connected, then the Kronecker graphs generated by such initiators are optimally searched in $\Theta(\sqrt{N})$ time. On the other hand, if the fixed, regular, and connected initiator is bipartite, then optimal quantum search is not achieved. In this bipartite case, although the runtime is slower than $O(\sqrt{N})$, it is still better than a classical computer's $O(N)$ runtime if $M > 2$.

In relation to prior results, our work is a generalization of \cite{Wong29}, which focused on the complete graph as the initiator, since the complete graph is an initiator with a dominant principal eigenvalue. While our results are based on a Lemma by \cite{Chakraborty2016}, their work applies it to Erd\H{o}s-R\'enyi random graphs, whereas we focus on Kronecker graphs. Both Erd\H{o}s-R\'enyi random graphs and Kronecker graphs are important in the study of complex networks, but the types of networks they generate have stark differences. Finally, our results differ from \cite{Chakraborty2018}, which used search Hamiltonian related to an interpolating Markov chain; our search Hamiltonian follows the original proposal of Childs and Goldstone \cite{CG2004}.

%-------------------------------------------------------------------------------

\section{\label{sec:dominant} Dominant Eigenvalue Initiators}

First, let us prove that if the initiator has a dominant principal eigenvalue, then the Kronecker graphs generated by it also have a dominant principal eigenvalue. We label and order the eigenvalues of the initiator graph's adjacency matrix $A$ as $\lambda_{A,1} \ge \lambda_{A,2} \ge \dots \ge \lambda_{A,M}$ with corresponding eigenvectors $\ket{v_{A,1}}, \ket{v_{A,2}}, \dots, \ket{v_{A,M}}$. Then, the eigenvalues of $A^{\otimes j}$, the adjacency matrix of the $j$th order Kronecker graph, are scalar products of the eigenvalues of $A$, and the eigenvectors of $A^{\otimes j}$ are the Kronecker products of the eigenvectors of $A$. As a proof,
\begin{align*}
	A^{\otimes j} & \left( \ket{v_{A,i}} \otimes \dots \otimes \ket{v_{A,k}} \right) \\
		&= (A \otimes \dots \otimes A) \left( \ket{v_{A,i}} \otimes \dots \otimes \ket{v_{A,k}} \right) \\
		&= A \ket{v_{A,i}} \otimes \dots \otimes A \ket{v_{A,k}} \\
		&= \lambda_{A,i} \ket{v_{A,i}} \otimes \dots \otimes \lambda_{A,k} \ket{v_{A,k}} \\
		&= \left( \lambda_{A,i} \dots \lambda_{A,k} \right) \left( \ket{v_{A,i}} \otimes \dots \otimes \ket{v_{A,k}} \right).
\end{align*}

Using this property of Kronecker products, it follows that the principal eigenvalue of $A^{\otimes j}$ is $(\lambda_{A,1})^j$. Similarly, the eigenvalue(s) of $A^{\otimes j}$ with the second-largest magnitude takes the form $(\lambda_{A,1})^{j-1} \lambda_{A,i}$ for some $i \in \{ 2, \dots, M \}$. For example, if the initiator is the complete graph of $M$ vertices, whose eigenvalues are given in \eqref{eq:complete}, then the $j$th order Kronecker graph has principal eigenvalue $(M-1)^j$, the second-largest eigenvalues in magnitude are $-(M-1)^{j-1}$ with multiplicity $M$, the third-largest eigenvalues in magnitude are $(M-1)^{j-2}$ with multiplicity $M(M-1)/2$, and so on. In general, the principal eigenvalue of $A^{\otimes j}$ asymptotically dominates all the other eigenvalues in magnitude since it contains at least one more factor of $\lambda_{A,1}$ than the other eigenvalues, and $\lambda_{A,1}$ dominates the other $\lambda_{A,{i \ge 2}}$. Therefore, if an initiator has a dominant principal eigenvalue, its Kronecker powers also have a dominant principal eigenvalue.

Now we prove that quantum search is fast on all Kronecker graphs whose initiators have a dominant principal eigenvalue. To do this, we utilize the following Lemma by Chakraborty \textit{et al.}~\cite{Chakraborty2016}, which we quote verbatim:
\begin{quote}
	\textbf{Lemma \cite{Chakraborty2016}:} Let $H_1$ be a Hamiltonian with eigenvalues $\lambda_1 \ge \lambda_2 \ge \dots \ge \lambda_k$ (satisfying $\lambda_1 = 1$ and $|\lambda_i| \le c < 1$ for all $i > 1$) and eigenvectors $\ket{v_1} = \ket{s}, \ket{v_2}, \dots, \ket{v_k}$ and let $H_2 = \ketbra{w}{w}$ with $\left| \braket{w}{s} \right| = \epsilon$. For an appropriate choice of $r = O(1)$, applying the Hamiltonian $(1+r)H_1 + H_2$ to the starting state $\ket{v_1} = \ket{s}$ for time $\Theta(1/\epsilon)$ results in a state $\ket{f}$ with $\left| \braket{w}{f} \right| \ge (1-c)/(1+c) - o(1)$.
\end{quote}
Note this Lemma uses a Hamiltonian that runs backward in time, c.f., \eqref{eq:Hplus}. As explained by Chakraborty \textit{et al.}~\cite{Chakraborty2016}, a quantum walk effected by the adjacency matrix can be connected to this Lemma by letting $H_1$ equal the adjacency matrix divided by its principal eigenvalue. For Kronecker graphs, if the eigenvalues of $A^{\otimes j}$ are labeled and ordered $\lambda_{A^{\otimes j}, 1} \ge \lambda_{A^{\otimes j}, 2} \ge \dots \ge \lambda_{A^{\otimes j}, N}$, we identify $H_1 = A^{\otimes j} / \lambda_{A^{\otimes j},1}$. Then $H_1$ has eigenvalues $\lambda_1 = 1$, $\lambda_2 = \lambda_{A^{\otimes j},2}/\lambda_{A^{\otimes j},1}$, and so on. Since $\lambda_{A^{\otimes j},1}$ dominates the other eigenvalues $\lambda_{A^{\otimes j},i \ge 2}$, each of the non-principal eigenvalues of $H_1$ limit to 0, asymptotically. So they are trivially bounded above in magnitude by a constant $c < 1$, satisfying the conditions of the Lemma. Then utilizing the Lemma, the system evolves to a state $\ket{f}$ that, upon measurement, yields the marked vertex with asymptotically constant probability, i.e., using big-$\Theta$ notation \cite{Knuth1976}, $|\braket{w}{f}|^2 = \Theta(1)$.

For the runtime, note a regular graph of $N$ vertices has the uniform state \eqref{eq:psi0} as its principal eigenvector, which is the initial state of the system \cite{Chung1997}. Then in the Lemma, $\epsilon = \braket{w}{s} = 1/\sqrt{N}$, so the runtime of the algorithm is $\Theta(1/\epsilon) = \Theta(\sqrt{N})$. Since the system reaches a constant success probability in time scaling as the square root of the number of vertices, Kronecker graphs with regular, dominant eigenvalue initiators are therefore optimally searched by a continuous-time quantum walk in $\Theta(\sqrt{N})$ time.

More specifically, the precise runtime of a single iteration of the algorithm can be deduced from Chakraborty \textit{et al.}'s \cite{Chakraborty2016} supplemental material. Combining (14) from their supplemental material with the sentence after their (19), the runtime is
\begin{equation}
	\label{eq:runtime}
	t_* = \frac{\pi}{2 \braket{v_1}{w}} \sqrt{ \sum_{i \ge 2} \frac{\braket{v_i}{w}^2}{(1-\lambda_i)^2}},
\end{equation}
where the phases of the $\ket{v_i}$'s are chosen so that their inner products with the marked vertex $\ket{w}$, i.e., $\braket{v_{i}}{w}$, are real and nonnegative.

\begin{figure*}
\begin{center}
	\subfloat[] {
		\label{fig:paley5}
		\includegraphics{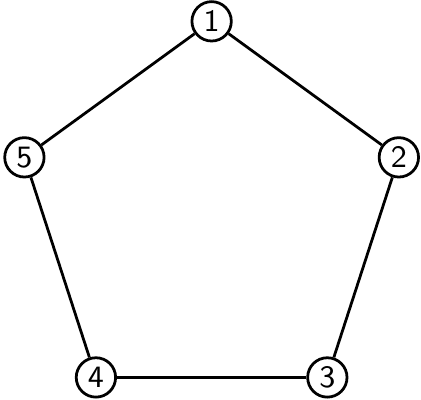}
		\quad \quad \quad
		\includegraphics{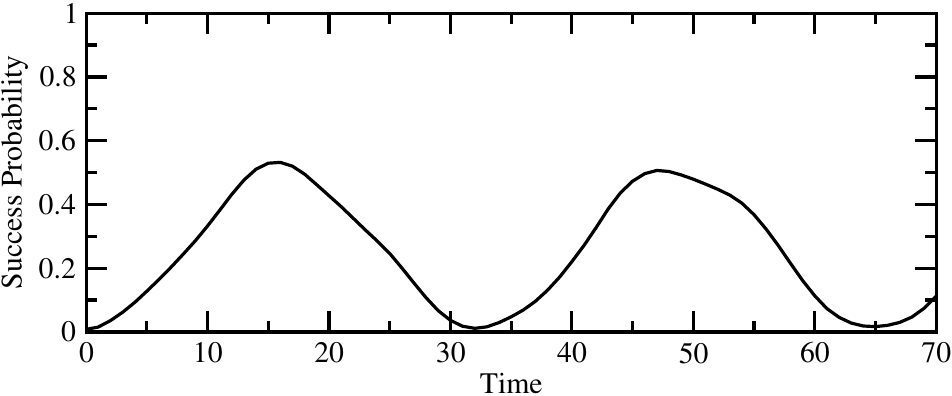}
	}

	\subfloat[] {
		\includegraphics{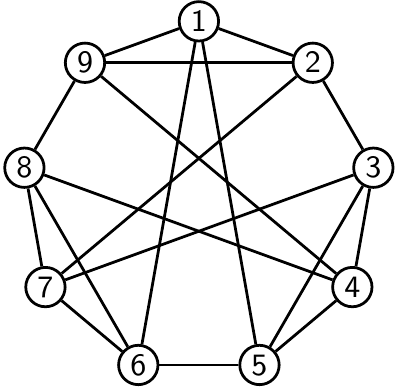}
		\quad \quad \quad
		\includegraphics{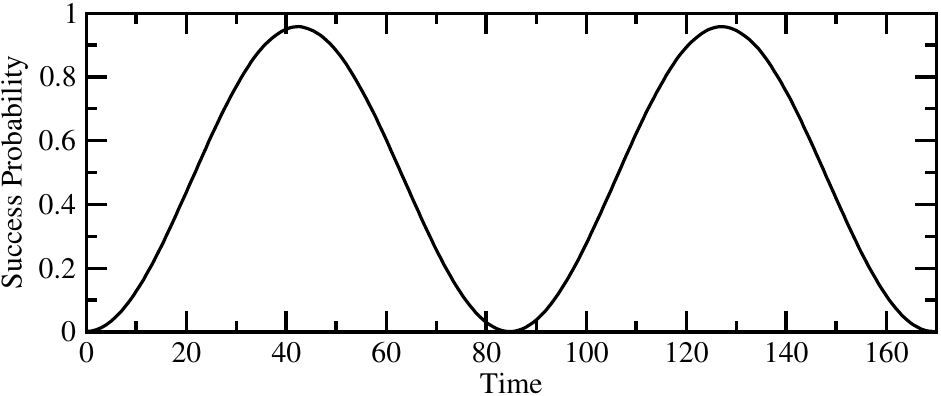}
	}

	\subfloat[] {
		\includegraphics{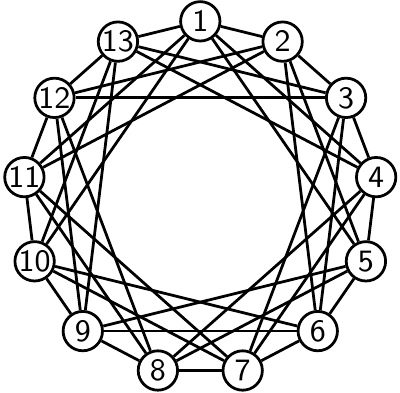}
		\quad \quad \quad
		\includegraphics{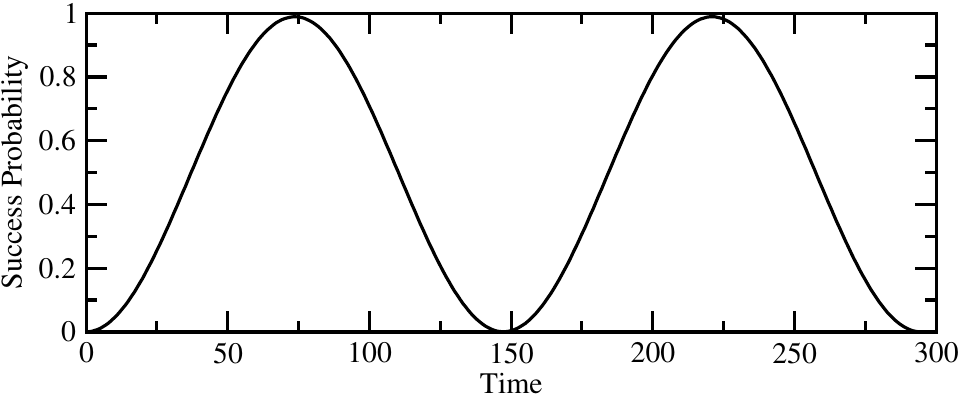}
	}
	\caption{\label{fig:paley_j3} Success probability as a function of time for search on the $j = 3$ order Kronecker graph generated by the Paley graph of (a) $M = 5$ vertices, (b) $M = 9$ vertices, and (c) $M = 13$ vertices with $\gamma = [2/(M-1)]^3$.}
\end{center}
\end{figure*}

Of course, this optimal search only occurs when the jumping rate $\gamma$ is chosen to take some critical value. To determine it, we can relate it to $r$ by comparing the search Hamiltonian in the Lemma with the typical backward-time search Hamiltonian \eqref{eq:Hplus}. Doing so, we identify
\begin{equation}
	\label{eq:gammar}
	\gamma = \frac{1+r}{\lambda_{A^{\otimes j},1}}.
\end{equation}
So the critical $\gamma$ can be determined from the critical $r$. Chakraborty \textit{et al.} \cite{Chakraborty2016} did not give an explicit closed-form solution for the critical $r$, but it can be deduced by combining (3) and (6) from their supplemental material, yielding
\begin{equation}
	\label{eq:rc}
	r_c = \frac{\sum_{i\geq 2} \frac{\braket{v_i}{w}^2}{1-\lambda_i} }{\sum_{i\geq 2}\braket{v_i}{w}^2} - 1.
\end{equation}
During the preparation of this manuscript, a similar, independent observation of $r_c$ was made by \cite{Chakraborty2018}. The inner products $\braket{v_i}{w}$ may be complicated to determine, so instead, we use a simpler approach given by Glos \textit{et al.}~\cite{Glos2018} for when $c$ scales less than a constant. Recall from the Lemma that $c$ is an upper bound on the magnitudes of the non-principal eigenvalues of $H_1$. Glos \textit{et al.}~\cite{Glos2018} noted that the supplemental material of \cite{Chakraborty2016} gives an explicit bound on $r$ between $-c/(1+c)$ and $c/(1-c)$, inclusive.  So if $c = o(1)$, then $r = o(1)$. This is true for Kronecker graphs with dominant eigenvalue initiators, since
\begin{align*}
	c 
		&= \frac{\max_{i = 2, \dots, N} |\lambda_{A^{\otimes j},i}|}{\lambda_{A^{\otimes j},1}} \\
		&= \frac{(\lambda_{A,1})^{j-1} \max_{i = 2, \dots, M} |\lambda_{A,i}|}{(\lambda_{A,1})^j} \\
		&= \frac{\max_{i = 2, \dots, M} |\lambda_{A,i}|}{\lambda_{A,1}} = o(1),
\end{align*}
where the last equality is because the initiator's principal eigenvalue dominates the others in magnitude. For example, if the initiator is the complete graph \eqref{eq:complete}, then $H_1 = A^{\otimes j}/\lambda_{A^{\otimes j},1}$ has eigenvalues $1$, $-1/(M-1)$, $1/(M-1)^2$, and so on, so the non-principal eigenvalues of $H_1$ are bounded above in magnitude by $c = 1/(M-1)$. Using $r = o(1)$ and \eqref{eq:gammar}, the asymptotic critical jumping rate for a $j$th order Kronecker graph with a dominant eigenvalue initiator is
\begin{equation}
	\label{eq:gammac}
	\gamma_c = \frac{1}{\lambda_{A^{\otimes j},1}} = \frac{1}{(\lambda_{A,1})^j}.
\end{equation}
For example, if the initiator is the complete graph \eqref{eq:complete}, then $\gamma_c = 1/(M-1)^j$, which is consistent with Wong \textit{et al.}'s \cite{Wong29} results when $j = 1,2,3$.

Glos \textit{et al.}~\cite{Glos2018} also noted that if $c = o(1)$, then the success probability is not just any constant, but is asymptotically $1$. Similarly, the runtime is not just $\Theta(\sqrt{N})$, but is asymptotically $\pi \sqrt{N}/2$. Since Kronecker graphs with dominant eigenvalue initiators satisfy $c = o(1)$, the success probability and runtime are
\begin{gather}
	p_* = 1, \label{eq:prob_dominant} \\
	t_* = \frac{\pi \sqrt{N}}{2}. \label{eq:runtime_dominant}
\end{gather}
Thus, we have proved that Kronecker graphs with dominant eigenvalue initiators are optimally searched by a continuous-time quantum walk, reaching a success probability of $1$ at time $\pi\sqrt{N}/2$, with the jumping rate chosen according to \eqref{eq:gammac}, for large $M$ and independently of $j$. Note $j = 1$ is the initiator graph itself, so any graph with a dominant eigenvalue, and its Kronecker powers, are optimally quantum searched. Since this result only depends on $M$, the Kronecker power $j$ can either be fixed or vary---its value is not important.

As a check of this analytical result, let us numerically simulate search on Kronecker graphs generated by Paley graphs. A Paley graph of $M$ vertices is defined when $M$ is a prime power congruent to $1 \pmod 4$, and vertices are adjacent if their distance is a square in the finite field ${\rm GF}(M)$ \cite{Cameron1991}. The Paley graphs with $M=5$, $9$, and $13$ vertices are shown in Fig.~\ref{fig:paley_j3}. The adjacency matrix of the Paley graph of $M$ vertices has a unique principal eigenvalue $(M-1)/2$, and the remaining eigenvalues are $(-\sqrt{M} - 1)/2$ and $(\sqrt{M} - 1)/2$ and are degenerate (see the supplemental material of \cite{Wong7} for more details). Thus, Paley graphs have a dominant principal eigenvalue, so our analytical results apply. We expect that for large $M$, the search algorithm \eqref{eq:H} with $\gamma$ chosen according to $\eqref{eq:gammac}$ should reach a success probability of $1$ \eqref{eq:prob_dominant} at time $\pi \sqrt{N} / 2$ \eqref{eq:runtime_dominant}. Numerically, this asymptotic behavior is confirmed in Fig.~\ref{fig:paley_j3} for third-order Kronecker graphs. When $M=5$, the Paley graph is too small for the asymptotic behavior to occur, but $M = 9$ is better, and $M = 13$ gives strong agreement, approaching a success probability of $1$ at time $\pi \sqrt{13^3} / 2 \approx 73.627$.

The main results from this section can be summarized by following theorem.
\begin{theorem}
	Let $A$ be an $M \times M$ adjacency matrix of a regular graph with principal eigenvalue $\lambda_{A,1}$, and $j$ a positive integer. If $\lambda_{A,1}$ dominates the other eigenvalues of $A$ for large $M$, then a quantum walk on $A^{\otimes j}$ with Hamiltonian $\gamma A^{\otimes j} + \ketbra{w}{w}$ and proper choice of jumping rate satisfying $\gamma = [1+o(1)]/\lambda_{A_1}^j$ evolves from the starting state $\ket{s}$ to a final state $\ket{f}$ with $|\braket{f}{w}|^2= 1-o(1)$ in time $\pi\sqrt{M^j}/2 + o(\sqrt{M^j})$.
\end{theorem}

Note that the Lemma by Chakraborty et al.~\cite{Chakraborty2016} only includes one marked vertex, as does our result. With multiple marked vertices, there may be many different spatial arrangement of the marked vertices, which could affect the jumping rate and runtime \cite{Wong9}. For example, for the complete graph, the arrangement of multiple marked vertices does not affect the behavior of the algorithm. But the Kronecker power of the complete graph is no longer complete, and two marked vertices could be adjacent or nonadjacent to each other, and this constitutes different cases that may require separate analysis. Hence, we leave multiple marked vertices as an open question.

%-------------------------------------------------------------------------------

\section{\label{sec:shift} Optimal Shifting and Rescaling of the Quantum Walk Hamiltonian}

Chakraborty \textit{et al}.~\cite{Chakraborty2016} noted that their Lemma implies that any regular graph with constant normalized algebraic connectivity supports optimal quantum search. Normalized algebraic connectivity is a measure of how connected a graph is, and a constant value indicates a relatively high level of connectedness. Their proof was limited to a footnote in their paper, however, so in this section, we provide a more thorough proof. In doing so, we show that any regular, non-bipartite graph with adjacency matrix $A$ and constant normalized algebraic connectivity has the property that the non-principal eigenvalues of $H_1 = A/\lambda_{A,1}$ are bounded in magnitude from $1$ by at least a constant. That is,
\begin{equation}
	\label{eq:H1bound}
	c_A = \max_{i \ge 2} \frac{|\lambda_{A,i}|}{\lambda_{A,1}} < 1.
\end{equation}
We then show that this is also true for regular, bipartite graphs, but it requires shifting and rescaling $H_1$. We derive optimal choices for the shifting and rescaling which minimize the upper bound $c$ in the Lemma, hence maximizing the lower bound on the success probability.

The algebraic connectivity of a graph is defined as the difference between the two smallest eigenvalues of the combinatorial Laplacian $L = D-A$, where $D_{ii} = \deg{(i})$ is the diagonal degree matrix \cite{Chung1997}. For a regular graph, each vertex has the same number of neighbors, so $D$ is a multiple of the identity matrix. Then $D$ does not change the difference between any eigenvalues---it only shifts all of them by a constant. Thus, for a regular graph, $D$ can be ignored, and the algebraic connectivity is the difference between the two smallest eigenvalues of $-A$, which is equal to the difference between the two largest eigenvalues of $A$, so it is $\lambda_{A,1} - \lambda_{A,2}$.

For a regular graph, the normalized algebraic connectivity is a rescaling of this difference; it is the algebraic connectivity divided by the degree of the graph \cite{Chung1997}. Since the degree of a regular, connected graph is equal to its principal adjacency eigenvalue $\lambda_{A,1}$, the normalized algebraic connectivity is $(\lambda_{A,1} - \lambda_{A,2})/\lambda_{A,1} = 1 - \lambda_{A,2}/\lambda_{A,1}.$ This is precisely the difference between the first two eigenvalues of $H_1 = A/\lambda_{A,1}$, i.e., $\lambda_1 - \lambda_2 = 1 - \lambda_2$. Furthermore, assuming the graph is connected, $\lambda_2$ must be less than $1$, so if the normalized algebraic connectivity is constant, then $\lambda_2$ is a constant less than $1$.

The normalized algebraic connectivity is also defined as the difference between the two smallest eigenvalues of the normalized combinatorial Laplacian $\mathcal{L}$, which for a regular graph is the combinatorial Laplacian $L$ divided by the degree of the graph. That is, $\mathcal{L} = L / \lambda_{A,1} = I - A/\lambda_{A,1} = I - H_1$. Then since the eigenvalues of the normalized Laplacian $\mathcal{L}$ are between $0$ and $2$, inclusive \cite{Chung1997}, the eigenvalues of $H_1$ are between $-1$ and $1$, inclusive. Thus, if the normalized algebraic connectivity is constant, then all $\lambda_{i \ge 2}$ are bounded from $\lambda_1 = 1$ by at least a constant.

Even though this proves that the non-principal eigenvalues are bounded away from $1$, it does not prove that they are bounded away from $-1$, which the Lemma requires. For example, for the regular complete bipartite graph $K_{n,n}$ (which has $2n$ vertices, $n$ in each partite set), its adjacency matrix has eigenvalues $n$, $0$, and $-n$ with respective multiplicities $1$, $2n-2$, and $1$. Dividing by the principal eigenvalue $n$, $H_1$ has eigenvalues $1$, $0$, and $-1$ with respective multiplicities $1$, $2n-2$, and $1$. Taking the difference between the two largest eigenvalues of $H_1$, the normalized algebraic connectivity is $1$, a constant. Yet the eigenvalue $-1$ has magnitude $1$, so the non-principal eigenvalues are upper-bounded in magnitude by $c = 1$, but the Lemma requires $c < 1$.

Although such a situation occurs if and only if the initiator is a bipartite graph, we can overcome this obstacle and prove that optimal quantum search still exists. For $K_{n,n}$, rather than using $H_1 = A/n$ to effect the quantum walk, consider instead $H_1' = (H_1 + 0.25 I) / 1.25$, for example. Other numbers can be used, but with these particular numbers, $H_1'$ has eigenvalues $1$, $0.2$, and $-0.6$ with respective multiplicities $1$, $2n-2$, and $1$. Thus, the non-principal eigenvalues are upper-bounded in magnitude by $c = 0.6$, and the Lemma implies optimal quantum search in $\Theta(\sqrt{N})$ time. Furthermore, since $0.25 I$ is a multiple of the identity matrix, it can be dropped, so $H_1' = H_1 / 1.25$. Since this is simply $H_1$ rescaled by a constant, if there exists optimal quantum search using $H_1'$ with some $r'$, then there exists optimal quantum search using $H_1$ with $r = (r' - 0.25)/1.25$. Note shifting and rescaling only affects the global phase and jumping rate.

We can always perform such a shift and rescaling if the normalized algebraic connectivity is constant because the eigenvalues of $H_1$ are between $-1$ and $1$, inclusive. So a constant normalized algebraic connectivity does imply optimal quantum search, as Chakraborty \textit{et al.}~\cite{Chakraborty2016} claimed.

We take this further by optimizing the shifting and rescaling of $H_1$ to maximize the success probability by minimize the upper bound $c$. If we consider $H_1' = (H_1+aI)/b$, then as proved in Appendix~\ref{sec:appendix}, the optimal choice of $a$ and $b$ are
\begin{gather*}
	a = - \frac{\lambda_2+\lambda_N}{2}, \\
	b = \frac{2-\lambda_2-\lambda_N}{2}.
\end{gather*}
With these values, the eigenvalues of $H_1'$ are
\begin{gather*}
	\lambda_1' = 1, \\
	\lambda_2' = \frac{\lambda_2-\lambda_N}{2-\lambda_2-\lambda_N}, \\
	\lambda_N' = -\lambda_2' = \frac{\lambda_N-\lambda_2}{2-\lambda_2-\lambda_N}.
\end{gather*}
Note that for $H_1'$, we have $c = \lambda_2'< 1$, which satisfies the condition of the Lemma.

Finally, note we can apply this same shifting and rescaling when the graph is non-bipartite in order to improve the lower bound $c$.

The main result from this section can be summarized by the following theorem:
\begin{theorem}
	Let $A$ be the adjacency matrix of a regular, connected graph of $N$ vertices with eigenvalues $\lambda_{A,1} > \lambda_{A,2} \ge \dots \ge \lambda_{A,N}$. Then
	\begin{gather*}
		a = - \frac{\lambda_{H_N,2}+\lambda_{H_N,N}}{2}, \\
		b = \frac{2-\lambda_{H_N,2}-\lambda_{H_N,N}}{2},
	\end{gather*}
	maximize Chakraborty et al.'s \cite{Chakraborty2016} lower bound on the success probability for quantum spatial search, where the quantum walk is effected by $H = (H_1 + aI)/b + \ketbra{w}{w}$ with $H_1 = A/\lambda_{A,1}$.
\end{theorem}

%-------------------------------------------------------------------------------

\section{\label{sec:fixed} Fixed Regular, Connected Initiators}

In Section~\ref{sec:dominant}, we explored search where $M$, the number of vertices in the initiator, was large. Here, we instead consider fixed initiators with constant $M$ and explore quantum search as the Kronecker power $j$ increases. We assume that the initiator is regular so that its principal eigenvector is the uniform superposition \eqref{eq:psi0}, and we also assume the initiator is connected so its principal eigenvalue is unique. Now let us consider separately when the initiator is non-bipartite or bipartite.

If the fixed regular, connected initiator is non-bipartite, we can prove that optimal $\Theta(\sqrt{N})$ quantum search occurs. First, since the $j$th Kronecker power of a regular initiator is regular, its principal eigenvector is the initial uniform superposition $\ket{s}$. Then in the Lemma, $\epsilon = \braket{w}{s} = 1/\sqrt{N}$, so the runtime of a single instance of the algorithm is $\Theta(1/\epsilon) = \Theta(\sqrt{N})$. Second, since the initiator graph is fixed, connected, and non-bipartite, it has constant normalized algebraic connectivity, and its adjacency eigenvalues satisfy \eqref{eq:H1bound} from the last section. Then for the $j$th order Kronecker graph generated by such an initiator, the non-principal eigenvalues of $H_1 = A^{\otimes j} / \lambda_{A^{\otimes j},1}$ are upper bounded in magnitude by
\begin{align*}
	c
		&= \frac{\max_{i=2,\ldots,N}|\lambda_{A^{\otimes j},i}|}{\lambda_{A^{\otimes j},1}} \\
		&= \frac{(\lambda_{A,1})^{j-1}\max_{i=2,\ldots,M}|\lambda_{A,i}|}{(\lambda_{A,1})^j} \\
		&=  \frac{\max_{i=2,\ldots,M}|\lambda_{A,i}|}{\lambda_{A,1}} \\
		&= c_A < 1,
\end{align*}
where the last line comes from \eqref{eq:H1bound}. Since this bound does not depend on $j$, the inequality holds for arbitrary, and hence large, $j$. Then from the Lemma, the success probability is $\Theta(1)$, and quantum search occurs in optimal $\Theta(\sqrt{N})$ time, for a fixed initiator and for large $j$. Of course, we can always lower the upper bound $c$, and hence increase the lower bound on the success probability, using the shifting and rescaling from the last section, but it will only change the constant factor, not the scaling of the algorithm.
This result can be summarized by the following theorem:
\begin{theorem}
	Let $A$ be an $M \times M$ adjacency matrix for a connected, regular, non-bipartite graph. Then there exists jumping rate $\gamma = \Theta(1/\lambda_{A}^j)$, such that walking by Hamiltonian $\gamma A^{\otimes j} + \ketbra{w}{w}$ for time $O(\sqrt{M^j})$ evolves the starting state $\ket{s}$ to a state $\ket{f}$ with $|\braket{f}{w}|^2= \Theta(1)$.
\end{theorem}

\begin{figure*}
\begin{center}
	\subfloat[] {
		\label{fig:paley_M5_j1}
		\includegraphics{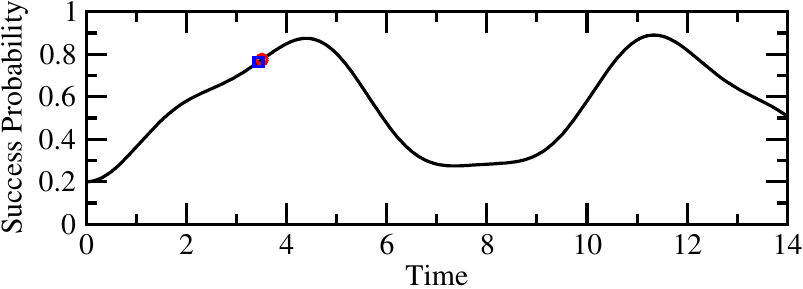}
	} \quad
	\subfloat[] {
		\label{fig:paley_M5_j2}
		\includegraphics{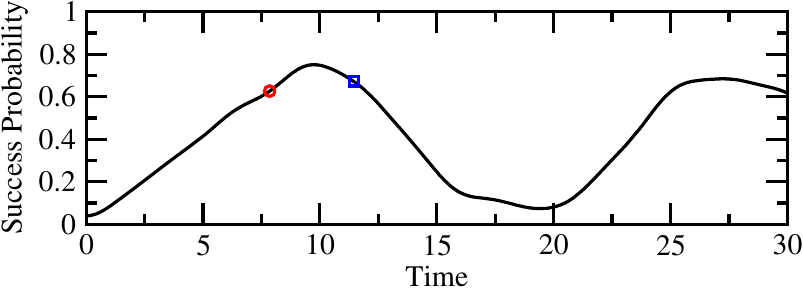}
	}

	\subfloat[] {
		\label{fig:paley_M5_j3}
		\includegraphics{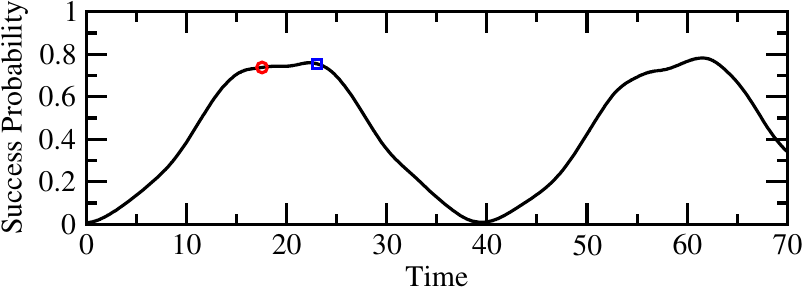}
	} \quad
	\subfloat[] {
		\label{fig:paley_M5_j4}
		\includegraphics{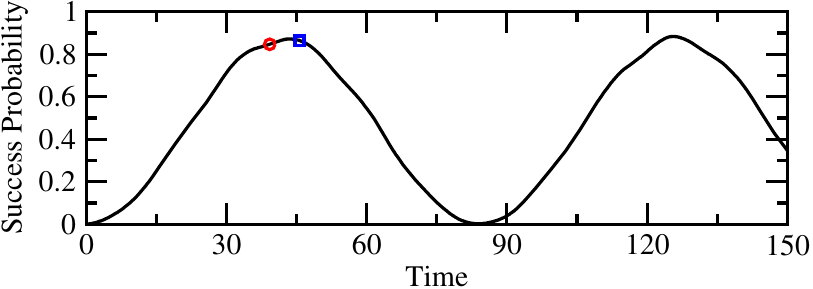}
	}

	\subfloat[] {
		\label{fig:paley_M5_j5}
		\includegraphics{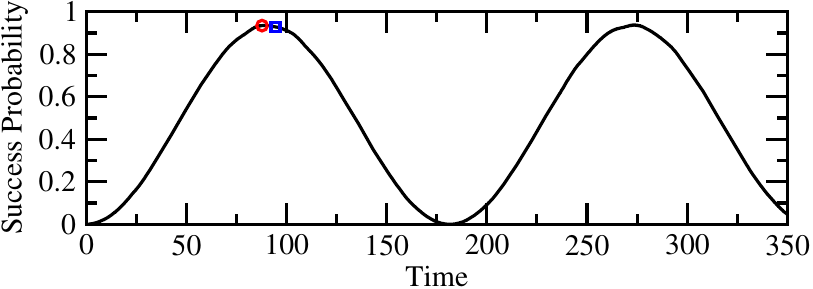}
	} \quad
	\subfloat[] {
		\label{fig:paley_M5_j6}
		\includegraphics{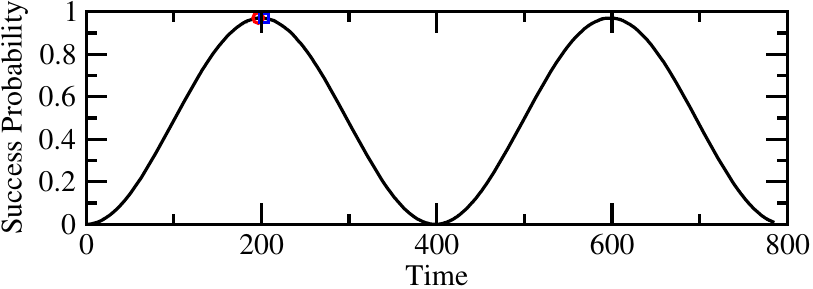}
	}

	\subfloat[] {
		\label{fig:paley_M5_j7}
		\includegraphics{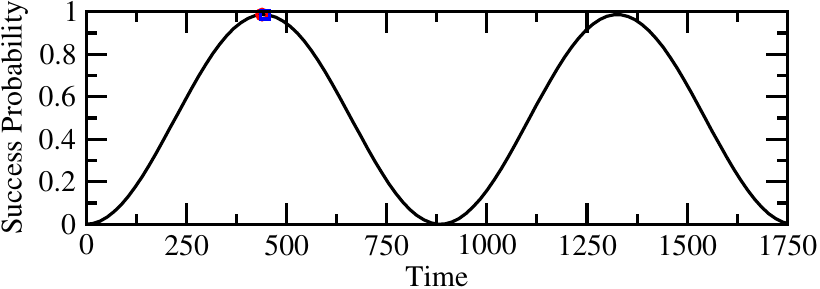}
	} \quad
	\subfloat[] {
		\label{fig:paley_M5_j8}
		\includegraphics{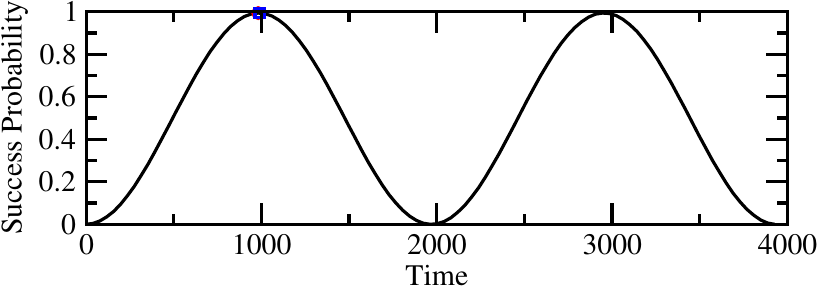}
	}
	\caption{\label{fig:paley_M5} Success probability as a function of time for search on the $j$th order Kronecker graph generated by the Paley graph of $M = 5$ vertices with $\gamma$ was chosen according to \eqref{eq:gammar} and \eqref{eq:rc}. (a) is $j = 1$, (b) is $j = 2$, (c) is $j = 3$, (d) is $j = 4$, (e) is $j = 5$, (f) is $j = 6$, (g) is $j = 7$, and (h) is $j = 8$. For each of these, the success probability at time $\pi \sqrt{N} / 2 = \pi \sqrt{5^j} / 2$ is indicated by a red circle, and as $j$ increases, it converges to a maximum success probability of $1$. Similarly, for each subfigure, the success probability at time $t_*$ \eqref{eq:runtime} is indicated by a blue square, and as $j$ increases, it converges to $\pi \sqrt{N}/2$.}
\end{center}
\end{figure*}

As a numerical check of this analytical result, consider the Paley graph of $M = 5$ vertices from Fig.~\ref{fig:paley5}, which is simply the cycle of $5$ vertices, as the initiator. The success probability as the system evolves with time is shown for the first eight Kronecker powers in Fig.~\ref{fig:paley_M5}, and $\gamma$ was chosen according to \eqref{eq:gammar} and \eqref{eq:rc}. On these plots, we also identified the success probability at time $\pi \sqrt{N} / 2$ as a red circle, and as $j$ increases, it converges to a maximum success probability of $1$. This is in agreement with our analytical results that the runtime is $\Theta(\sqrt{N})$ with a success probability of $\Theta(1)$. It is known that cycles are quantum searched in classical $O(N)$ time \cite{CG2004}, yet Kronecker powers of a fixed cycle are quantum searched in optimal $O(\sqrt{N})$ time. Note for the cycle graph of 5 vertices, we have $c=1/\sqrt{5}$, and so the success probability is asymptotically lower bounded by $(1-c)/(1+c) = (3-\sqrt{5})/2 \approx 0.382$. In this example, the optimal success probability is much larger than its lower bound. The subfigures also include a blue square showing the success probability at time $t_*$ \eqref{eq:runtime} that was derived from the Lemma, and it converges to $\pi\sqrt{N}/2$ for large $N$.

The situation differs if the initiator graph is bipartite. In this case, we have $\lambda_{A,M}=-\lambda_{A,1}$, and hence the second-order Kronecker graph always consists of at least two principal eigenvalues equal to $(\lambda_{A,1})^2$. This prevents direct use of the Lemma, which requires that the principal eigenvalue be unique. Still, we can construct a graph with optimal quantum search if we allow shifting and rescaling the initiator's adjacency matrix. For example, for the complete bipartite graph, we can instead use $A' = (A + 0.25 I) / 1.25$. In this way, we reduce the problem to the previously considered, non-bipartite scenario. Since the shifting and rescaling introduces self-loops and weighted edges, this is not quite the same graph, however, and its Kronecker powers also differ from the unshifted graph.

\begin{figure}
\begin{center}
	\subfloat[] {
		\includegraphics{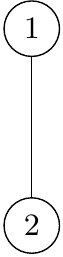}
	} \quad \quad \quad \quad \quad
	\subfloat[] {
		\includegraphics{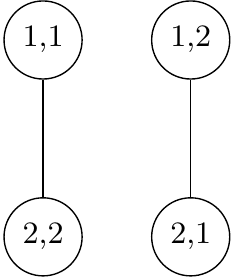}
	}

	\subfloat[] {
		\includegraphics{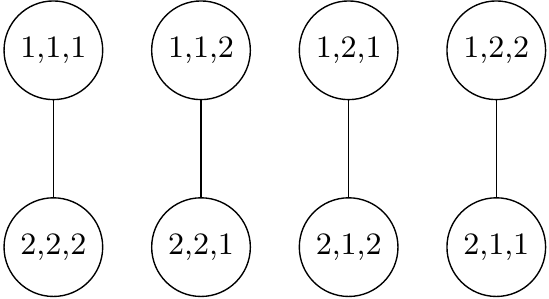}
	}
	\caption{\label{fig:P2powers}(a) The path graph of two vertices $P_2$, (2) its second-order Kronecker graph $P_2 \otimes P_2$, and (c) its third-order Kronecker graph $P_2 \otimes P_2 \otimes P_2$.}
\end{center}
\end{figure}

A natural alternative would be to shift $A^{\otimes j}$ instead of $A$. This does not yield optimal quantum search, however, as we now prove. First, note that if a bipartite graph is $d$-regular with partite sets $V_1$ and $V_2$, then the number of vertices in each set must be the same, i.e., $|V_1|=|V_2|$. This comes from the fact that the sums of the degrees of each partite set must be equal, since edges only exist between the sets. For a $d$-regular bipartite graph, this implies $d|V_1| = d|V_2|$ or $|V_1| = |V_2|$. Then, if the initiator is bipartite and regular, each of its partite sets contains $M/2$ vertices.

Next, the Kronecker graphs generated by a bipartite initiator consist of bipartite graphs that are disconnected from each other. An example is shown in Fig.~\ref{fig:P2powers}, where the initiator is the path graph of two vertices, which is bipartite, and its second- and third-order Kronecker powers consist of two and four separate bipartite graphs, respectively. In general, if $V_1$ and $V_2$ are the partite sets of the initiator, then the second-order Kronecker graph consists of two components with respective vertex sets $(V_1 \times V_2) \cup (V_2 \times V_1)$ and $(V_1 \times V_1) \cup (V_2 \times V_2)$, where $\times$ denotes the Cartesian product of sets. Since $|V_1| = |V_2| = M/2$, $|V_1 \times V_1| = |V_1 \times V_2| = |V_2 \times V_1| = |V_2 \times V_2| = (M/2)^2$, so each component has $2(M/2)^2$ vertices. Generalizing this, since the Kronecker product acts on disconnected components independently, the $j$th Kronecker power of a regular bipartite graph results in $2^{j-1}$ regular, bipartite graphs that are disconnected from each other, each with $M^j/2^{j-1} = 2\left( M/2 \right)^j$ vertices, for a total of $N = M^j$ vertices.

To determine the runtime of the quantum walk search algorithm with bipartite initiators, we again start with the $M=2$ case in Fig.~\ref{fig:P2powers} in order to build intuition. Then we will generalize to arbitrary $M$. If the initiator has $M=2$ vertices, then it is the path graph of order 2, and the Kronecker power is a collection of path graphs of the same order, see again Fig.~\ref{fig:P2powers}. Since the evolution of quantum spatial search on disconnected components is independent, the success probability can grow no larger than $2/N$, so the total runtime with classical repetitions is $\Theta(N)$. Thus, we do not achieve optimal quantum search. Furthermore, it is the same complexity as classically, randomly guessing for a marked vertex.

Now for general $M$, which must be even since $|V_1| = |V_2| = M/2$ implies there is no regular, connected, bipartite graph of odd vertices, we similarly prove that optimal quantum search is not obtained because the Kronecker product produces multiple components that are disconnected from each other, restricting the total success probability. To elucidate the proof, we begin with the $j=2$ case, and then we generalize it to arbitrary $j$. Since the eigenvalues of the adjacency matrix of a bipartite graph are symmetric \cite{BH2012} (i.e., $\lambda_{A,1} = -\lambda_{A,M}$, $\lambda_{A,2} = -\lambda_{A,M-1}$, and so forth), the largest, second largest, and smallest eigenvalues of $A^{\otimes 2}$  take the form $\lambda_{A,1}^2$, $\lambda_{A,1} \lambda_{A,2}$, and $-\lambda_{A,1}^2$, with respective multiplicities 2, at least 4 (since $\lambda_{A,2}$ may not be unique), and 2. Since there are two largest eigenvalues, two disconnected components, and each component has a unique largest eigenvalue, each must correspond to a different component. Since the components are both bipartite, each component must have unique $\lambda_{A,1}^2$ and $-\lambda_{A,1}^2$ eigenvalues. Hence, the eigenvalue gap is $\lambda_{A,1}^2-\lambda_{A,2}\lambda_{A,1}$, and after dividing by $\lambda_{A,1}^2$, the normalized algebraic connectivity is equal to the $j=1$ case, and is hence constant.

Now for arbitrary $j$, the $j$th order Kronecker power has principal eigenvalue $\lambda_{A,1}^j$ with multiplicity $2^{j-1}$. Since we have $2^{j-1}$ bipartite graphs, each with $M^j/2^{j-1}$ vertices, that are disconnected from each other, each $\lambda_{A,1}^j$ corresponds to one bipartite graph. Furthermore, the second-largest eigenvalue for each component is at most $\lambda_{A,1}^{j-1} \lambda_{A,2}$. So each bipartite graph has an eigenvalue gap of at least $\lambda_{A,1}^j - \lambda_{A,1}^{j-1} \lambda_{A,2}$. Dividing by $\lambda_{A,1}^j$, the algebraic connectivity of each bipartite component is $1 - \lambda_{A,2} / \lambda_{A,1}$, which is constant (independent of $j$) for fixed $M$. Utilizing the shift and rescaling from Section~\ref{sec:shift}, the time it takes for the success probability to build up at the marked vertex is $O(\sqrt{M^j/2^{j-1}})$. Hence, if we start the evolution in a state equally spanned by vertices from the component with the marked element, thanks to the Lemma and the shifting and rescaling from Section~\ref{sec:shift}, after time $t_* = \Theta(\sqrt{ M^j / 2^{j-1}})$ we can find the marked element with probability $\Theta(1)$. The initial amplitude, however, is distributed evenly between all connected components, so the actual success probability is $p_* = 1/2^{j-1}$. Overall, the expected total runtime with classical repetitions is $t_*/p_*$, which scales as
\[ \frac{\sqrt{ \frac{M^j}{2^{j-1}}}}{\frac{1}{2^{j-1}}} = \frac{\pi}{\sqrt 2}\sqrt{2^jM^j} = \Theta \left( N^{\frac{1}{2}+\frac{1}{2\log_2 M}} \right), \] 
where the last equality comes from $N=M^j$, which implies $j = \log_2N/\log_2M$, and in turn $2^j =N^{1/\log_2 M}$. Note for large $j$ and arbitrary, constant $M$, this total runtime is greater than $\Theta(\sqrt{N})$, so optimal quantum walk search is not achieved. It is better, however, than a classical computer's runtime of $\Theta(N)$ when $M > 2$. When $M=2$, this formula yields our previous result of a runtime of $\Theta(N)$. Finally, note these results depend on $M$ alone and not the specific form of the initiator, other than it being regular, connected, and bipartite.

These results concerning connected, regular, bipartite initiators can be summarized in the following theorem:
\begin{theorem}
	Let $A$ be an $M \times M$ adjacency matrix for a connected, regular, bipartite graph. Then there exists jumping rate $\gamma = \Theta(1/\lambda_{A}^j)$, such that walking by Hamiltonian $\gamma A^{\otimes j} + \ketbra{w}{w}$ for time $t = O(\sqrt{M^j/2^{j-1}})$ evolves the starting state $\ket{s}$ to a state $\ket{f}$ with $p = |\braket{f}{w}|^2 = \Theta(1/2^{j-1})$. Then the expected runtime with classical repetitions is
	\[ t/p = \Theta \left( N^{\frac{1}{2}+\frac{1}{2\log_2 M}} \right). \]
\end{theorem}

%-------------------------------------------------------------------------------

\section{Conclusion}

Kronecker graphs are used in network science to generate complex networks with the characteristics of real-world networks. We proved that any Kronecker power of a regular graph that has a unique principal eigenvalue that asymptotically dominates the other eigenvalues can be optimally searched by a continuous-time quantum walk, reaching a success probability of $1$ at time $\pi \sqrt{N}/2$, asymptotically, with the jumping rate chosen according to \eqref{eq:gammac}. For example, the complete graph, Paley graph, and others have dominant principal eigenvalues. Not only do they asypmotically support optimal $O(\sqrt{N})$ quantum search, but any Kronecker power of them also supports optimal quantum search.

Furthermore, if the initiator is fixed, then we proved that taking successive Kronecker powers of the initiator yields graphs that are optimally searched if the initiator is regular, connected, and non-bipartite. For example, cycles by themselves are quantum searched in classical $O(N)$ time, but the Kronecker graphs generated by a cycle do support optimal quantum search in $O(\sqrt{N})$ time.

If the fixed, regular initiator is bipartite, however, then the quantum walk does search more quickly than a classical computer when the initiator has more than two vertices, although optimal quantum search is not achieved. When the initiator has two vertices, then it searches with the same scaling as a classical computer.

Altogether, these results greatly expand our knowledge of how quantum computers search Kronecker graphs. 

%-------------------------------------------------------------------------------
% Acknowledgments.
%-------------------------------------------------------------------------------

\begin{acknowledgments}
	T.W.~was partially supported by startup funds from Creighton University.
\end{acknowledgments}

%-------------------------------------------------------------------------------
% Appendix
%-------------------------------------------------------------------------------

\appendix

\section{\label{sec:appendix} Optimal Shifting and Rescaling}

As shown in Sec.~\ref{sec:fixed}, the quantum walk Hamiltonian $H_1$ can be shifted and rescaled to $H_1' = (H_1 + aI)/b$ in order to maximize the lower bound on the success probability in the limit. Here we will show that the values of $a$ and $b$ proposed in Sec.~\ref{sec:fixed} are the optimal ones.

First, recall that in the Lemma, the success amplitude is lower bounded in magnitude by $(1-c)/(1+c)$. The derivative of this is $-2/(1+c)^2$, which means the bound is a decreasing function on $c \in [0,1]$. Hence the maximization of the success amplitude or success probability is equivalent to minimizing $c$ on $[0,1]$.

Let us consider $H_1' = (H_1+aI)/b$ for some $a$ and $b$. Recall $H_1$ has eigenvalues $\lambda_1 = 1 \ge \lambda_2 \ge \dots \ge \lambda_N \ge -1$. Assuming the graph is connected, $\lambda_2 < 1$. Furthermore, since the sum of the eigenvalues of $A$ is $\text{tr}(A) = 0$ \cite{BH2012}, the sum of the eigenvalues of $H_1 = A/\lambda_{A,1}$ is also $0$, so we also have $\lambda_N < 0$. Similarly, recall $H_1'$ has eigenvalues $\lambda_1' = 1 \ge \lambda_2' \ge \dots \ge \lambda_N'$. Then the upper bound on the magnitudes of the non-principal eigenvalues is given by
\[ c = \max(|\lambda_2'|,|\lambda_N'|). \]

We consider $a$ and $b$ such that $\lambda_1' = 1$, which is a requirement for the Lemma. This implies
\[ \lambda_1' = \frac{\lambda_1+a}{b} = \frac{1+a}{b} = 1, \]
and so $b=1+a$. Substituting this, we have $H_1' = (H_1'+aI)/(1+a)$. Then,
\[ \lambda_2' = \frac{\lambda_2+a}{1+a}, \]
\[ \lambda_N' = \frac{\lambda_N+a}{1+a}. \]
Note that $|\lambda_2'| < 1$ and $|\lambda_N'| < 1$ need to be satisfied for the Lemma, otherwise the upper bound on their magnitude would be $c = 1$. Focusing on $|\lambda_2'| < 1$, this implies $a > -(1+\lambda_2)/2$, and since $\lambda_2 \le 1$, we have $a > -1$.

Now substituting our expressions for $\lambda_2'$ and $\lambda_N'$ into the upper bound $c$, we get
\[ c(a) = \frac{1}{|1+a|} \max(|\lambda_2+a|, |\lambda_N+a|). \]
The task is to find the value of $a > -1$ that minimizes $c$. We do this by considering three cases, when $|\lambda_2 + a|$ dominates the maximum, when $|\lambda_N + a|$ dominates the maximum, and when they are equal.

First, when $a$ is greater than the threshold value
\[ a_{\rm thr} = -\frac{\lambda_2+\lambda_N}{2}, \]
then $|\lambda_2+a|$ is dominant. Hence,
\[ c(a) = \left| \frac{\lambda_2+a}{1+a} \right|. \]
Since  $a > a_{\rm thr} > -\lambda_2> -1$, where the last inequality comes from $\lambda_2 < 1$ by assuming the graph is connected, we can drop the absolute value and write
\[ c(a) = \frac{\lambda_2+a}{1+a} = 1+ \frac{\lambda_2-1}{1+a}. \]
For $a > -1$, this function increases as $a$ increases. Thus, the minimum of $c$ occurs when $a \le a_{\rm thr}$.

Next, when $a$ is less than the threshold $a_{\rm thr}$, then $|\lambda_N+a|$ is dominant, and we have
\[ c(a) = \left| \frac{\lambda_N+a}{1+a} \right|. \]
Since $a < a_{\rm thr}$, we have $\lambda_N+a<\lambda_N - \frac{\lambda_2+\lambda_N}{2} = \frac{\lambda_N-\lambda_2}{2} \le 0$, where the last inequality comes from $\lambda_2 \ge \lambda_N$. Thus,
\[ c(a) = -\frac{\lambda_N+a}{1+a} = -1 +\frac{1-\lambda_N}{1+a}. \]
For $a > -1$, this function decreases as $a$ increases. Thus, the minimum of $c$ occurs when $a \ge a_{\rm thr}$.

Combining these results proves that $c$ is minimized when $a$ equals its threshold value, at which $c$ is
\[ c_{\rm min} = \frac{|\lambda_2-\lambda_N|}{|2-\lambda_2-\lambda_N|}. \]
Using $a = a_{\rm thr}$, with $b = 1+a$, $\lambda_2' = (\lambda_2 + a)/(1 + a)$, and $\lambda_N' = (\lambda_N + a)/(1 + a)$, we get the values stated in Sec.~\ref{sec:shift}.

%-------------------------------------------------------------------------------
% References.
%-------------------------------------------------------------------------------

\bibliography{refs}

\end{document}